\documentclass[12pt]{article}

\font\msytw=msbm10 scaled\magstep1%
\font\msytww=msbm8 scaled\magstep1%

\let\a=\alpha \let\b=\beta  \let\g=\gamma  \let\d=\delta \let\e=\varepsilon
  \let\h=\eta   \let\th=\vartheta \let\k=\kappa \let\l=\lambda
\let\m=\mu    \let\n=\nu    \let\x=\xi     \let\p=\pi    \let\r=\rho
 \let\t=\tau    
     
\let\G=\Gamma    
         
\let\O=\Omega 

\def\Eq#1{{\label{#1}}%
}
\def\equ#1{(\ref{#1})}

\newdimen\xshift \newdimen\xwidth \newdimen\yshift \newdimen\ywidth

\def\ins#1#2#3{\vbox to0pt{\kern-#2pt\hbox{\kern#1pt #3}\vss}\nointerlineskip}

\def\eqfig#1#2#3#4#5{
\par\xwidth=#1pt \xshift=\hsize \advance\xshift
by-\xwidth \divide\xshift by 2
\yshift=#2pt \divide\yshift by 2
{\hglue\xshift \vbox to #2pt{\vfil
#3 \includegraphics{#4.eps}
}\hfill\raise\yshift\hbox{#5}}}

\def\eqalign#1{\null\,\vcenter{\openup\jot
  \ialign{\strut\hfil$\displaystyle{##}$&$\displaystyle{{}##}$\hfil
      \crcr#1\crcr}}\,}

\usepackage{fancyhdr}
\def\lis#1{\overline#1}
\def\wt#1{\widetilde#1}

\def\ZZZ{\hbox{\msytw Z}}
\def\QT{\hbox{\msytw T}} 
\def\QZ{\hbox{\msytw Z}} \def\Qzz{\hbox{\msytww Z}}

\def\QN{\hbox{\msytw N}}

\let\0=\noindent
\def\*{\vskip3mm}
\def\defi{\,{\buildrel def\over=}\,}
\def\Val{{\rm Val}}
\def\AA{{\cal A}}
\def\GG{{\cal G}}

\def\PP{{\cal P}}
\def\TT{{\cal T}}

\def\bea{\begin{eqnarray}}\def\eea{\end{eqnarray}}
\let\nn=\nonumber
\def\be{\begin{equation}}\def\ee{\end{equation}}

\def\iniz{\setcounter{equation}{0}{
\rhead{\thepage}\lhead{{{{\small\bf\thesection:}\ \SEC}}}}}

\begin{document}

\centerline{\Large\bf Resonances within Chaos}

\vskip3mm

\centerline{G. Gallavotti${}^{1,2}$, G. Gentile${}^3$, A. Giuliani${}^3$}
\vskip.1truecm
\small{\centerline{${}^1$ Dip.to di Fisica, 
Univ. di Roma ``La Sapienza", 00185 Roma - Italy}}
\vskip.1truecm
\small{\centerline{${}^2$ Dept. of Mathematics, Rutgers University, NJ 08854-8019 USA}}
\vskip.1truecm
\small{\centerline{${}^3$ Dip.to di Matematica, 
Univ. di Roma Tre, 00146 Roma - Italy}}
\vskip.2truecm

{\vskip3mm}
\noindent {\bf Abstract}: {\it A chaotic system under periodic forcing can
  develop a periodically visited strange attractor. We discuss simple
  models in which the phenomenon, quite easy to see in numerical
  simulations, can be completely studied analytically.}  {\vskip3mm}

{\bf In chaotic systems synchronization phenomena, i.e. resonances,
  can occur: the simplest being motions asymptotically developing over a
  low dimension attractor which is visited in synchrony with a weak
  time-periodic perturbing force. This is analogous to the synchronization
  (``phase locking'') occurring when an integrable system, perturbed by a
  periodic forcing, develops attracting periodic orbits with period being a
  simple fraction (1, 3/2, 2,\ldots) of the forcing period.  Formation
  of an attractor and existence of dissipation, i.e. of phase volume
  contraction, are concomitant. A chaotic system under a weak volume
  preserving periodic perturbation can have, instead, a very singular
  chaotic evolution, \cite{SW000,RW001}, with dense orbits. A simple
  illustration of the resonance phenomenon in a chaotic system is exhibited
  here and exactly solved by deriving the shape of the attractor: which
  turns out to be a continuous surface with weak smoothness properties
  (being a strange attractor with an exponent of H\"older-continuity with a
  very small lower bound).
  The peculiarity of the class of systems treated is the
  possibility of determining analytically an attractor whose existence can
  be rather easily seen in simulations.  }

\*
\def\SEC{Resonances and chaos}
\section{\SEC}\label{sec1}\iniz

A chaotic system weakly interacting with a very regular system may generate
a fully chaotic dynamics: as a basic example, consider the suspension flow
of Arnold's cat map $A_2=\pmatrix{2&1\cr1&1\cr}$
with a flat floor function (i.e. the flow on $\QT^3= [0,2\p]^3$ where
every point $x\in\QT^2$ of the base moves vertically with constant velocity
until it reaches the flat floor, at which point it reappears on the base, at
the location $A_2x$) coupled with a clock, which moves at the same speed as
the suspension flow. If observed at the times when the suspension flow hits
the floor, the unperturbed dynamics takes the form of a linear discrete
dynamical system $A_3: \QT^3\to\QT^3$, with $A_3\defi
\pmatrix{2&1&0\cr1&1&0\cr0&0&1\cr}$. 
Consider now a small perturbation of
$A_3$: it is remarkable that in the case of volume preserving
  perturbations, some pathological phenomena can take place
  \cite{SW000,RW001}. Namely, there is an open set $U$ of $C^1$
  conservative perturbations of $A_3$ such that the following happens: the
  perturbed motion $g\in U$ is ergodic with respect to the volume measure
  in $\QT^3$ and there exists a foliation of $\QT^3$ into $C^2$ curves
  $\ell_x$ that are invariant under $g$ (i.e. $g\ell_x=\ell_{g x}$);
  moreover, the central Lyapunov exponent along the leaves is positive on
  the set $\O$ of full three-dimensional measure where it is
  defined; the set $\O$  intersects every leaf at exactly $k$ points, for
  some $k>0$ \cite{SW000,RW001}.
Such results provide an interesting example of an absolutely singular 
foliation, in the sense of Katok \cite{Mi997}. 

The situation is destined to change substantially in the presence of
friction: if we couple via a dissipative perturbation the suspension flow
of $A_2$ with a clock, the phase between the two (that is arbitrary in the
unperturbed case) might lock, making the motion fall onto a lower
dimensional attractor, which would naturally be close to the union of $k$
elements of a ``horizontal'' foliation of $\QT^3$.  We can say, when this
happens, that there is a resonance between the periodic forcing and the
chaotic evolution on $\QT^3$: and unlike the cases quite typical in generic
perturbations of periodically forced integrable systems where the resonance
may lead to the appearance of stable periodic orbits, in the presence of
hyperbolicity and of small dissipation the forcing gives rise to a
``periodic strange attractor'', i.e. to an attractor for the time evolution
observed at times integer multiples of the forcing period, on which the
dynamics is conjugate to a uniformly hyperbolic system.

In this paper we provide examples of the above phenomenon: remark that in
the non-dissipative case the perturbations discussed in \cite{SW000,RW001}
can be in a set arbitrarily close to the unperturbed system and open in the
$C^1$ topology (not containing the unperturbed system itself). Here we
consider a class of analytic perturbations depending on a parameter $\e$
and show that for any small $\e\ne0$ the resonance occurs.  The phenomenon
is reminiscent of certain synchronization phenomena in chaotic systems
\cite{BKOVZ002} and the simple case considered below is, as far as we know,
one of the first explicit examples of such phenomena that can be fully
worked out analytically.

\def\SEC{A simple example}
\section{\SEC}
\label{sec2}\iniz

Denoting by $x=(x_1,x_2)$ a point of $\QT^2$,
consider the evolution equation:
\be\eqalign{
\dot x & = \d(z)\,(Sx-x) + \e \, f(x,z,w) , \cr 
\dot w & = 1 + \e \, g(x,z,w) ,\cr
\dot z & = 1 , \cr 
}
\Eq{e2.1} \ee
with $(x,w,z)\in \QT^2 \times \QT \times \QT$ and $\e\ge0$. Here $\d$ is the
$2\p$-periodic delta function, with the convention that
$\int_{0}^{t}{\rm d}t'\,\d(t')=1$ for all $t>0$. Set $f=0$ to simplify as
much as possible the problem. The map $S$ could be any linear hyperbolic
automorphism of $\QT^2$, in particular $S=A_2$.

Then we can prove the following result.

\*\0{\bf Theorem:} {\it Let $S_g^t$ be the flow on $\QT^4$
associated with Eq.\equ{e2.1} with $f=0$ and $g$ analytic. Let also
\be \lis \g_0(x,w) \defi \int_{0}^{2\p} {\rm d}t \, g(Sx,t,w+t) , \qquad
\lis\g_1(x,w)\defi\int_0^{2\p} 
{\rm d}t \, \frac{\partial}{\partial w} g(Sx,t,w+t) , \Eq{e2.2}\ee
and suppose existence of $w_0$ such that $\lis \g_0(x,w_0)=0$ and
$\lis\g_1(x,w_0)=\G<0$, independent of $x,w$.
Then there are constants $\e_0,C>0$ such that for $0<\e<\e_0$ 
there exists a H\"older-continuous function
$W:\QT^2\to\QT$ of exponent $\b\ge C\e$  and
  
\begin{itemize}
\item the surface $\AA\defi\{(x,W(x),0): x\in\QT^2\}$ is invariant
under the Poincar\'e map $S_g^{2\p}$; 
\item the dynamics of $S_g^{2\p}$ on $\AA$ is 
conjugated to that of $S$ on $\QT^2$, i.e.
\be S_g^{2\p}(x,W(x),0)=(Sx,W(Sx),0)\;.\Eq{e2.3}\ee
\end{itemize}
}
\*

{\bf Remarks.}\\ (1) It should be clear from the proof below that, for
$\e>0$, $\AA$ is an attractor for $S_g^t$ with negative central Lyapunov
exponent equal to $\frac{\e\G}{2\p}(1+O(\e))$, see concluding remarks in
Section \ref{sec7}.  Furthermore the function $W$ is holomorphic in $\e=\r
e^{i\a}$ inside the domain which in polar coordinates is
$\r(\a)=\e_0\cos\a$.  
\\ 
(2) A simple explicit example of a function $g$
satisfying the assumptions of the theorem with $w_0=\p$ and $\G=-1$ is:
\be g(x,z,w) = \sin \left( w-z \right) +
\sin \left( x_{1} + w + z \right) .
\Eq{e2.4} \ee
(3) It would be interesting to estimate the positions of the poles of the
power spectrum of the correlations of pairs of smooth observables in the
sense of the chaotic resonances theory in \cite{Ru986}. The spectrum is
meromorphic in a strip with poles off the real axis for a dense set of
observables: the unperturbed system has poles at $-i\log\l_++2\p k$,
$k\in \ZZZ$, if $\l_+$ is the largest eigenvalue of the matrix $S$; we
expect the position of the poles closest to the origin to be
modified by order $\e$. Notice that in the unperturbed system the spectrum
is an entire function for all analytic observables and the poles
really appear with non-zero residue for H\"older continuous functions
(see Sec. 4.3 and Fig.4.3.5 in \cite{GBG004}).\\

Let us now turn to the proof of the theorem. The main ideas and the 
strategy of the proof are the following: 
(i) after having written the equation for the invariant manifold, we solve it 
recursively; (ii) the result of the recursion is conveniently expressed in 
terms of tree diagrams; (iii) these can be easily shown to give 
rise to a convergent expansion, provided that the large factors $\e^{-1}$ 
produced under iterations by the small rate of contraction of phase space 
(of order $O(\e)$) is compensated by suitable cancellations, stemming 
from the condition that $\lis\g_0(x,w_0)=0$. All these steps are discussed in 
detail in the following Sections \ref{sec3}--\ref{sec6}. Finally, in Section
\ref{sec7}, we add some comments on the assumptions made and on possible 
generalizations of our result. 

\def\SEC{The equation for the invariant manifold}
\section{\SEC}\label{sec3}\iniz

Let $W(x)=w_0+U(x)$, $x(0)=x$ and for $0<t\le 2\p$ write 
$w(t)=w_0+t+u(x,t)$, with 
$u(x,0^+)=U(x)$. Substituting into Eq.\equ{e2.1}, 
for all $0<t\le 2\p$ we find:
\be \dot u(x,t) = \e \, g(Sx,w_{0}+t+u(x,t),t) .
\Eq{e3.1} \ee
Defining 
\be \eqalign{ & \g_{0}(x,t) = g(Sx,w_{0}+t,t) , \qquad \g_{1}(x,t) =
{\partial \over \partial w} g(Sx,w_{0}+t,t) , \cr 
& G(x,t,u) = g(Sx,w_{0}+t+u,t)
- \g_{0}(x,t) - u \,\g_{1}(x,t) , \cr} \Eq{e3.2} \ee
Eq.\equ{e3.1} can be rewritten for all 
$0 <t \le2\pi$ as
\be \dot u(x,t) = \e u(x,t)\,\g_{1}(x,t) + \e \Big(
\g_{0}(x,t) + G(x,t,u(x,t)) \Big) ,
\Eq{e3.3} \ee
with
\be G(x,t,u)=\sum_{p\ge 2} G_p(x,t)
  u^p\;. \Eq{e3.4} \ee
Integrating from $0$ to $t\in(0,2\p]$, the equation for $u(x,t)$ becomes
\be u(x,t) = {\rm e}^{\e\G(x,t,0)} U(x) + \e \int_{0}^{t} {\rm d}\t \,
{\rm e}^{\e\G(x,t,\t)} \Big( \g_{0}(x,\t) + G(x,\t,u(x,\t)) \Big) ,
\Eq{e3.5} \ee
where
\be \G(x,t,\t) = \int_{\t}^{t} {\rm d}\t' \, \g_{1}(x,\t') .
\Eq{e3.6} \ee
By construction $G$ vanishes to second order in $u$, hence, if $\d$
is fixed once and for all, one has
$ u(x,2\pi) = {\rm e}^{\G\e} u(x,0) + O(\e^{2})$, so that
\be |u(x,2\pi)| \le {\rm e}^{\G\e/2} |u(x,0)| , \qquad
\hbox{if } \, {\d\over2} \le |u(x,0)| \le \d ,
\Eq{e3.7} \ee
provided $\e$ is small enough. Therefore the region $|u|\le \d$
is an attracting set for $\e$ small (compared to $\d$).

The condition of invariance for $W(x)$ reads $u(x,2\pi)=U(Sx)$, which gives
\be U(Sx) - {\rm e}^{\e\G} U(x) =
\e \int_{0}^{2\pi} {\rm d}\t \,
{\rm e}^{\e\G(x,2\pi,\t)} \Big( \g_{0}(x,\t) + G(x,\t,u(x,\t)) \Big) ,
\Eq{e3.8} \ee
and hence
\be \eqalign{
U(x) =& \sum_{k=1}^{\infty} {\rm e}^{\e\G (k-1)} F(S^{-k}x) ,
\cr
F(x) =& \e \int_{0}^{2\pi} {\rm d}\t \,
{\rm e}^{\e\G(x,2\pi,\t)} \Big( \g_{0}(x,\t) + G(x,\t,u(x,\t)) \Big) .\cr}
\Eq{e3.9} \ee
In order to solve the two coupled equations Eqs.\equ{e3.5}-\equ{e3.8}, we find
convenient to consider the two-parameters $(\m,\e)$ equations
\be  \eqalign{
& \hskip-.2truecm
u(x,t) = {\rm e}^{\m\G(x,t,0)} U(x) + 
\e \int_{0}^{t} {\rm d}\t \,
{\rm e}^{\m\G(x,t,\t)} \Big( \g_{0}(x,\t) + G(x,\t,u(x,\t)) \Big) ,
\cr
& \hskip-.2truecm 
U(Sx) - {\rm e}^{\m\G} U(x) =
\e \int_{0}^{2\pi} {\rm d}\t \,
{\rm e}^{\m\G(x,2\pi,\t)} \Big( \g_{0}(x,\t) + G(x,\t,u(x,\t)) \Big)\;,\cr} 
\Eq{3.2}\ee
first showing existence of a solution analytic in $\e$ and
H\"older-continuous in $x$, at fixed $\m$, and subsequently proving that
$\m$ can be taken to have values $\m \ge \e$ a little beyond $\e$.

\def\SEC{Recursion equations}
\section{\SEC}\label{sec4}\iniz

Eq.(\ref{3.2}) can be solved recursively as follows. Let 

\be u(x,t)= {\rm e}^{\m \G(x,t,0)} U(x) +\x(x,t)\Eq{e4.1}\ee
and 
write $\x(x,t)=\e \x_{1}(x,t)+\e^{2}\x_{2}(x,t) + \cdots$ and
$U(x)=\e U_{1}(x) +\e^{2} U_{2}(x)+\cdots$.
Plugging the expansions into Eq.(\ref{3.2}) and setting $\l={\rm
  e}^{\m \G}$, we find
\be \eqalign{
& \x_{1}(x,t)  = \int_{0}^{t} {\rm d}\t \, {\rm e}^{\m\G(x,t,\t)}
\g_0(x,\t) , \qquad \x_{2}(x,t) = 0\;,\nn \cr
& U_{1}(x)  = \sum_{k=1}^{\infty} \l^{k-1}
\int_{0}^{2\p} {\rm d}\t \, {\rm e}^{\m\G(S^{-k}x,2\p,\t)}
\g_{0}(S^{-k} x,\t) , \qquad
U_{2}(x)  = 0, \cr}
\Eq{e4.2} \ee
and, for $n\ge 3$,
\bea &&\hskip-.7truecm\x_n(x,t)=
\int_{0}^{t} {\rm d}\t \sum_{p=2}^{\infty}\hskip-.2truecm
\sum_{n_{1},\ldots,n_{p} \ge 1 \atop n_{1}+\ldots+n_{p}=n-1}
\kern-4mm
{\rm e}^{\m \G(x,t,\t)} G_{p}(x,\t) \,\prod_{i=1}^p\Big(
 {\rm e}^{\m \G(x,\t,0)} U_{n_i}(x) +\x_{n_i}(x,t)\Big) , \nn\\
&&\hskip-.7truecm U_n(Sx)-{\rm e}^{\m\G}U_n(x)=
\int_{0}^{2\p} {\rm d}\t \sum_{p=2}^{\infty}\hskip-.2truecm
\sum_{n_{1},\ldots,n_{p} \ge 1 \atop n_{1}+\ldots+n_{p}=n-1}
\kern-4mm
{\rm e}^{\m \G(x,2\p,\t)} G_{p}(x,\t) \cdot\Eq{e4.3}
\\
&&\hskip4.5truecm\cdot\prod_{i=1}^p\Big(
 {\rm e}^{\m \G(x,\t,0)} U_{n_i}(x) +\x_{n_i}(x,t)\Big) ,
\nn\eea
which give a recursive definition of $\x_n,U_n$. Note that the second of 
Eq.\equ{e4.3} can be solved in a way analogous to Eq.\equ{e3.9}, so that
\be\eqalign{
& U_{n}(x) = \sum_{k=1}^{\infty} \l^{k-1}
\int_{0}^{2\p} {\rm d}\t \sum_{p=2}^{\infty}\sum_{n_{1},\ldots,n_{p}
     \ge 1 \atop n_{1}+\ldots+n_{p}=n-1}
{\rm e}^{\m \G(S^{-k} x,2\p,\t)} G_{p}(S^{-k} x,\t)  \, \cdot \cr
&\hskip3.truecm\cdot \prod_{i=1}^p
\Big( {\rm e}^{\m\G(S^{-k}x,\t,0)}U_{n_i}(S^{-k}x,\t)+
\x_{n_i}(S^{-k}x,\t)\Big) .
\cr}\Eq{e4.4}\ee
For the incoming discussion, it is useful to note that by the analyticity
of $g$, there exists a constant $C_0$ such that
\be |G_p(x,t)|\le C_0^p\;,\qquad {\rm e}^{\m\G(x,t,\t)}\le C_0\;,
\Eq{bounds}\ee
uniformly in $x,t,\t$. Moreover, if 
we consider the Fourier expansion
\be\eqalign{G_p(x,\t)=& \!\! \sum_{\n\in\Qzz^{2}}
{\rm e}^{i\n \cdot x}\hat G_p(\n,\t) ,\qquad
{\rm e}^{\m\G(x,t,\t)}= \!\! \sum_{\n\in\Qzz^{2}}
{\rm e}^{i\n \cdot x}\wt\G(\n,t,\t),\cr}
\Eq{e4.6}\ee
there exists $\k>0$, such that
\be |\hat G_p(\n,\t)|\le C_0 {\rm e}^{-\k |\n|},\qquad 
|\wt \G(\n,t,\t)|\le C_0 {\rm e}^{-\k |\n|} ,\Eq{e4.7}
\ee
uniformly in $t,\t$.

\def\SEC{Tree expansion}
\section{\SEC}\label{sec5}\iniz

The recursion for $\x_n,U_n$ can be conveniently represented graphically in
terms of tree graphs. We start with some basic definitions.
A connected graph $\GG$ is a collection of points (nodes)
and lines connecting all of them. We denote with
$V(\GG)$ and $L(\GG)$ the set of nodes and the set of lines,
respectively. A path between two nodes is the minimal subset of
$L(\GG)$ connecting the two nodes. A graph is planar if it
can be drawn in a plane without graph lines crossing.

A tree is a planar graph $\GG$ containing no closed loops.  One can
consider a tree $\GG$ with a single special node $v_{0}$: this introduces a
natural partial ordering on the set of lines and nodes, and one can imagine
that each line carries an arrow pointing toward the node $v_{0}$ thus
establishing a partial order $\succ$ on the tree.  We can add an extra
(oriented) line $\ell_{0}$ exiting the special node $v_{0}$; the added line
$\ell_{0}$ will be called the root line and the point $r$ it enters ({\it
  which is not a node}) will be called the root of the tree. In this way we
obtain a rooted tree $\theta$ defined by $V(\theta)=V(\GG)$ and
$L(\theta)=L(\GG)\cup\ell_{0}$. 

Hence given two nodes $v,w\in V(\theta)$, it is $v \prec w$ (or $w\succ v$)
if $w$ is on the path connecting $v$ to the root line.

A labeled tree is a rooted tree $\theta$
together with a label function defined on the sets $L(\theta)$ and
$V(\theta)$.

We shall call equivalent two rooted trees which can be transformed
into each other by continuously deforming the lines in the plane
in such a way that the latter do not cross each other
(i.e. without destroying the graph structure).
We can extend the notion of equivalence also to labeled trees,
simply by considering equivalent two labeled trees if they
can be transformed into each other in such a way that also
the labels match.

We can identify a line
with the nodes it connects; given a line $\ell=wv$ we say that
$\ell$ enters $w$ and exits $v$, and we shall write also $\ell=\ell_{v}$.
Given
two comparable nodes $v$ and $w$, with $v \prec w$, we denote with
$\PP(v,w)$ the path of lines connecting $w$ to
$v$, with $v$ and $w$ being included.  For any node $v \prec v_{0}$ we
denote by $v' \succ v$ the node immediately following $v$, hence
$\ell_{v}=v'v$, and we set $v_{0}'=r$.

Associate with the nodes and lines of any tree $\theta$ some labels,
according to the following rules.

With each node $v$ we associate the labels $(\h_{v}, k_{v})$, where
$\h_{v}=0,1$, while $k_{v}=0$ if $\h_{v}=0$ and $k_{v}\in\QN$
(i.e. $k_v\ge 1$) if $\h_{v}=1$; we call $\h_{v}$ the type label of the node
$v$.
Furthermore we define
\be k(v) = \sum_{v \in \PP(v,v_{0})} k_{v} .
\Eq{e5.1} \ee
For each node there are $p_{v} \ge 0$ entering lines.
If $p_{v}=0$ we say that $v$ is an end-node. If $p_{v}>0$ then $p_{v} \ge 2$.
For $p_{v} \ge 2$ let $s_{v}$ be the number of nodes $w$ with $w'=v$ such that
$\h_{w}=1$.

Represent graphically the node $v$ as a big white ball
(white node) if $\h_{v}=1$ and as a small black ball
(black node) if $\h_{v}=0$. Thus $s_{v}$ is
the number of white nodes immediately preceding $v$.

With each node $v$ we associate a time variable $\t_{v} \in (0,2\pi]$, with
$\t_{v_0'}=t$, a node factor $F_{v}=F_v(x,\t_{v'},\t_{v})$ defined as
\be  \null \hskip-.3truecm F_v=\cases{
{\rm e}^{\m\G(S^{-k(v)}x,\t_{v'},\t_v)} \,
G_{p_v}(S^{-k(v)}x,\t_{v}) \,
{\rm e}^{\m s_v \G(S^{-k(v)}x,\t_v,0)} , & $\h_{v} = 0$, \cr
\l^{k_v-1} {\rm e}^{\m\G(S^{-k(v)}x,2\p,\t_v)}\,
G_{p_v}(S^{-k(v)}x,\t_{v}) \,
{\rm e}^{\m s_v \G(S^{-k(v)}x,\t_v,0)} , & $\h_{v} = 1$, \cr}
\Eq{e5.2} \ee
and an integral $I_{v}(\h_{v})$, with
\be
I_{v}(0) = \int_{0}^{\t_{v'}} {\rm d}\t_{v},
\qquad I_{v}(1) = \int_{0}^{2\p} {\rm d}\t_{v}.
\Eq{e5.3} \ee

With the definitions above, we denote by $\TT_{n,\n,\h}$ the set of trees with
$n$ nodes and $\h_{v_{0}}=\h$. Then one can check (for instance by induction) 
that
\bea && U_{n}(x) = \sum_{\theta\in\TT_{n,1}} {\rm Val} (x;\theta) ,
\qquad \x_{n}(x,t) = \sum_{\theta\in\TT_{n,0}} {\rm Val} (x;\theta) ,\nn\\
&&
{\rm Val}(x;\theta) =
\prod_{v \in V(x;\theta)} I_{v}(\h_{v}) F_{v}(x,\t_{v'},\t_{v}) ,
\Eq{e5.4} \eea
where the integrals are understood to be performed by following
the tree ordering (i.e. by starting from the end-nodes and moving toward the
root). Note that in the formula for $\x_n(x,t)$ in Eq.\equ{e5.4},
the variable $\t_{v_{0}'}=t$ plays no role, as
$I_{v_{0}}(1)$ and $F_{v_0}(t,\t_{v_{0}})$ in fact do not depend on $t$.

For instance $\hat\x_{1}(\n,t)$ and $\hat U_{1}(\n)$ are
represented graphically as in Figure 1.

\*
\eqfig{214}{15}{
\ins{-52}{13.5}{$\x_{1}(x,t) =$}
\ins{097}{13.5}{$U_{1}(x) =$}
}{figure1}{}
\centerline{{\small {\bf Figure 1.} Graphical representation of
$\x_{1}(x,t)$ and $U_{1}(x)$.}}
\*

Since $\x_2,U_2=0$ the next simplest example is the representation 
of $\x_3,U_3$.
For instance, say, $U_{3}(x)$ is described in Figure 2.

\*
\eqfig{316}{55}{
\ins{-47}{34.5}{$U_{3}(x) =$}
\ins{70}{34.3}{$+$}
\ins{150}{34.3}{$+$}
\ins{240}{34.3}{$+$}
}{figure2}{}
\centerline{{\small {\bf Figure 2.} Graphical representation of $U_{3}(x)$.}}
\*

In a similar way we can represent graphically the other $U_n$'s or $\x_n$'s.
For example in Figure 3 we represent some contributions to $U_{5}(x)$:
all the other contributions are obtained by replacing
some white nodes with black nodes and vice versa, and possibly permuting
the lines entering the nodes (the only node which must be necessarily 
white is the node $v_{0}$).

\*
\eqfig{290}{100}{}{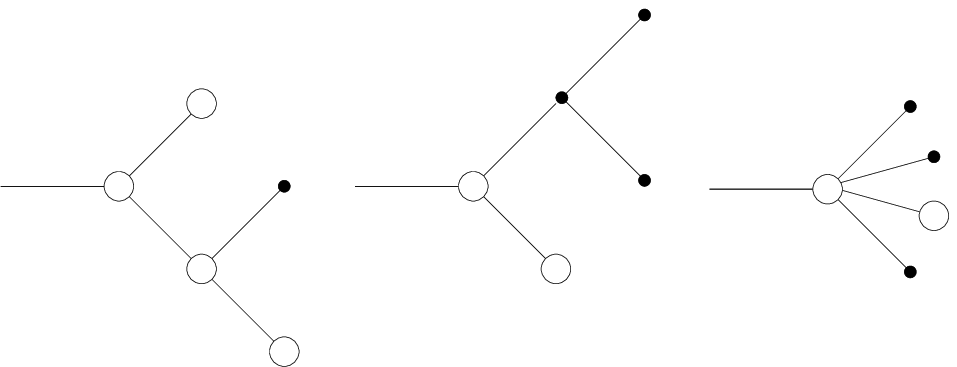}{}
\*
\centerline{{\small {\bf Figure 3.} Graphical representation of
some contributions to $U_{5}(x)$.}}
\*

In drawing the graphs in Figures 1--3 we have used that $p_{v}\ge 2$
whenever $p_{v} \neq0$ and that $\x_{2}(x,t)$ and $U_2(x)$ vanish
identically, as it follows from \equ{e4.2}.

\def\SEC{Convergence}
\section{\SEC}\label{sec6}\iniz

Given a tree $\th\in \TT_{n,x,\h}$, $\h\in\{0,1\}$, let $E(\th)$ be the
number of its end-nodes, and let $E_0(\th)$ and $E_1(\th)$ be the number of
end-nodes of type $0$ and $1$, respectively. Similarly, let $N^i(\th)$ be
the number of ``internal'' vertices of $\th$ (i.e. the number of vertices
of $\th$ with $p_v>0$) and $N^i_\h(\th)$ be the number of internal vertices
of type $\h$, with $\h=0,1$.  Note that, since $p_{v}\ge 2$ whenever $p_{v}
\neq0$, we have $E(\th)\ge (n+1)/2$.  A key remark is that, thanks to the
condition that $\lis\g_0(x,w_0)=0$, the contribution associated to every
end-node $v$ of type $1$ is of order $\m$,
\be \Big|\int_0^{2\p}{\rm d}\t_v \l^{k_v-1} 
{\rm e}^{\m\G (S^{-k(v)}x,2\p,\t_v)}\,
\g_0(S^{-k(v)}x,\t_v)\Big|\le ({\rm const}.)\m. \Eq{e6.1}\ee
Using this cancellation and the bounds Eq.\equ{bounds}, 
we find that, for all $n\in\QN$ and all $\n\in\QZ^{2}$,
\be \left| {\rm Val}(x;\theta) \right| \le C^{n}_{1}\m^{E_1(\th)}
\prod_{v \in V(\theta) \atop \h_{v}=1} \l^{k_{v}-1} ,
\Eq{e6.2} \ee
for a suitable constant $C_{1}$, uniformly in $x$.

Therefore, in order to obtain a bound on $U_{n}(x)$
we have to estimate the number of labeled trees in $\TT_{n,1}$.
For each node $v$ with $\h_{v}=1$ the corresponding sum over $k_{v}$
produces a factor $C_{2}\m^{-1}$. Finally the number of unlabelled
trees with $n$ nodes is bounded by $C_{3}^{n}$, with $C_{3}=4$.

By collecting together the bounds above we find
\be \left| U_{n}(x) \right|
\le C^{n}\m^{- N^i_1(\th)} , \qquad C = C_{1}C_{2}C_{3} .
\Eq{e6.3} \ee
Using $N^i_1(\th)\le N^i(\th)=n-E(\th)\le n-(n+1)/2$, we find
\be \left| U_{n}(x) \right|  \le C^{n}\m^{-(n-1)/2}\;,\ee
which yields a radius of convergence in $\e$ given
by $\e_{0}= C^{-1}\m^{1/2}$, hence we can take $\m=\e$.

To study the regularity in $x$ we need to get a similar upper bound 
for 
\be
\frac{|U_{n}(x)-U_{n}(x')|}{|x-x'|^{\b}} , \qquad \b> 0 .
\Eq{e6.5} \ee
We proceed as follows.
Given a tree $\th$ we introduce an arbitrary ordering of the nodes by setting
$V(\th)=\{v_{1},\ldots,v_{|V(\th)|}\}$ and define
$V^{-}_{p}(\th)=\{v_{1},\ldots,v_{p}\}$ and
$V_{p}^{+}=\{v_{p+1},\ldots,v_{|V(\th)|}\}$,
with the convention that
$V^{-}_{0}(\th)=V^{+}_{|V(\th)|}(\th)=\emptyset$ and
$V^{-}_{|V(\th)|}(\th)=V^{+}_{0}(\th)=V(\th)$. Then we define
\be \Val_{p}^{\pm} (x;\th) = \prod_{v \in V_{p}^{\pm}(\th)}
I_{v}(\h_{v}) F_{v}(x,\t_{v'},\t_{v}) , \Eq{e6.6}\ee
with $\Val_{0}^{-}(x;\th)=\Val_{|V(\th)|}^{+}(x;\th)=1$, and write
\be \eqalign{
& U_{n}(x)-U_{n}(x') = \sum_{\th\in\TT_{n,1}}
\Big(  \Val(x;\th) - \Val(x';\th) \Big)  \cr
& \qquad = \sum_{\th\in\TT_{n,1}} \sum_{p=1}^{|V(\th)|}
\Val_{p-1}^{-}(x;\th) \, \Val_{p+1}^{+}(x';\th) \, \cdot \cr
& \qquad \cdot \left( I_{v_{p}}(\h_{v_{p}})
\Big( F_{v_{p}}(x,\t_{v_{p}'},\t_{v_{p}}) -F_{v_{p}}(x',\t_{v_{p}'},
\t_{v_{p}}) \Big) \right) .
\Eq{e6.7} \cr} \ee
Each summand in \equ{e6.7} can be studied as in the case $\b=0$,
with the difference that the node factor is
$F_{v}(x,\t_{v'},\t_{v})$ for $v\in V_{p-1}^{-}(\th)$,
$F_{v}(x',\t_{v'},\t_{v})$ for $v\in V_{p+1}^{+}(\th)$ and
$F_{v}(x,\t_{v'},\t_{v}) - F_{v}(x',\t_{v'},\t_{v})$ for $v=v_{p}$.
Using Eqs.\equ{e5.2} and \equ{e4.6}, we find that 
\be \eqalign{
&  F_{v}(x,\t_{v'},\t_{v})=\sum_{\n\in\Qzz^{2}}
{\rm e}^{i\n \cdot x}\hat F_v(\n,\t_v,\t_{v'})\;,\cr
& |\hat F_v(\n,\t_v,\t_{v'})|\le \l^{(k_v-1) \d_{\h_v,1}}
\bar C_{1} \bar C_0^{s_v+2} {\rm e}^{-\frac{\k}{2}|S^{k(v)}\n|}\;,
\Eq{e6.8} \cr} \ee
for suitable constants $\bar C_0,\bar C_1$, independent of
$\t_v,\t_{v'}$ (in deriving the bound on the Fourier coefficients we
used that $F_{v}(x,\t_{v'},\t_{v})$ depends on $x$ through
$S^{-k(v)}x$, see Eq.\equ{e5.2}). Therefore,
\be \eqalign{
& \frac{|F_{v}(x,\t_{v'},\t_{v}) -F_{v}(x',\t_{v'},\t_{v})|}{|x-x'|^{\b}} \cr
& \hskip3.truecm \le \l^{(k_v-1) \d_{\h_v,1}}
\bar C_1 \bar C_0^{s_v+2}\sum_{\n\in\Qzz^{2}}|\n|^{\b} 
{\rm e}^{-\frac{\k}{2}|\n|}
|\l_-|^{-\b k(v)}\;,\Eq{e6.9} \cr} \ee
where $\l_-$ is the eigenvalue of the hyperbolic automorphism $S$ with
smallest absolute value; note that, if $\l_+$ is the other eigenvalue of
$S$, we have $|\l_-|=|\l_+|^{-1}$, with $|\l_+|>1$. The factor $|\l_-|^{-\b
  k(v)}$ in Eq.\equ{e6.9} is bounded by $\prod_{v\in V(\th):
  \h_v=1}|\l_+|^{\b k_v}$, so that
\be  \frac{|U_{n}(x)-U_{n}(x')|}{|x-x'|^{\b}} \le 
\sum_{\th\in\TT_{n,1}}n\bar C^n 
\m^{E_1(\th)}  \prod_{v\in V(\th)\atop
\h_{v}=1} \l^{k_{v}-1} |\l_{+}|^{\b k_{v}} ,\Eq{e6.10}\ee
for a suitable constant $\bar C$. We must require that $|\l_{+}|^{\b}\l<1$,
and in order for the sum $\sum_{k_v\ge 1}(|\l_{+}|^{\b}\l)^{k_v}$ to be
still proportional to $\m^{-1}$, we can take
\be \b = c \m , \qquad 0< c < |\G|/\log|\l_+| ,
\Eq{e6.11} \ee
which gives $\b= c\,\e$ for $\m=\e$. This concludes the 
proof of the theorem. 

\def\SEC{Concluding remarks}
\section{\SEC}\label{sec7}\iniz

Let us add a few comments on the role of the specific hypotheses in the theorem. 
The assumption that there is $w_0$ such that $\int_0^{2\p} {\rm
  d}t \, g(x,t,w_0+t)$ is zero has been heavily used as well as the
assumption that $\int_0^{2\p} {\rm d}t \, \partial_w g(x,t,w_0+t) =\G$, for
$\G<0$.  Both assumptions can be relaxed into

\begin{description}
\item{(a)}  $\int_0^{2\p} {\rm d}t \, g(x,w_0+t,t) = \e \lis g(x)$, for some
$\lis g(x)$,

\item{(b)} $\int_0^{2\p} {\rm d}t \, \partial_w g(x,w_0+t,t)  \le \G$, 
for $\G<0$,
\end{description}

\noindent and the statement of the theorem remains valid (and its proof is
essentially unchanged).

Assumption (a) is strong and it does not allow for perturbations of the form
$\e g$ because the $\e$-dependence has to involve higher powers of $\e$ to
satisfy (a) unless $\lis g(x)\equiv 0$ (as in the theorem proved here).

It is unclear whether the results discussed in this paper would 
hold under the weaker assumptions that

\begin{description}
\item{(a$'$)} $\int_0^{2\p} {\rm d}t \, g(x,w_0+t,t) = \wt g(x)$, 
with $\wt g(x)$
with $0$ average,

\item{(b$'$)} $\int_0^{2\p} {\rm d}t \, \partial_w g(x,w_0+t,t) =\wt
  g_1(x)$, 
with
  $\wt g_1(x)$ with negative average.

\end{description}

Preliminary computations and numerical evidence suggest that the existence of 
an attractor $\AA=\{(x,0,W(x))\}$ for $S_g^{2\p}$ may still be true, possibly
``just'' for a.e.-$x$ and with a non-smooth function $W(x)$: in other words,
a similar phase-locking phenomenon as the one proved in this paper may still 
take place, with $\AA$ a fractal rather than a smooth surface. 

It would be interesting to check that the central Lyapunov exponent in
the theorem is also expressed by an expansion in tree graphs: the procedure
should be the same illustrated in the theory of the tangent map for
perturbations of the simple Anosov map (e.g., $A_2$ in
Section \ref{sec1}), see proposition 10.3.1 in \cite{GBG004}.

Finally a natural conjecture is that the assumption $f=0$ in Eq.\equ{e2.1}
can be eliminated as long as $f$ is analytic. However, we do not have specific
indications of what is the general structure of the attractor in this case.
\*

{\bf{Acknowledgments.}} 
We gratefully acknowledge financial support from the Grant
``Sistemi dinamici, equazioni alle derivate parziali e meccanica statistica
(2008R7SKH2)" (G.G., G.G. and A.G.) and from ERC 
Starting Grant CoMBoS-239694 (A.G.). A.G. thanks the Institute for Advanced Studies for 
hospitality during the completion of this work.
\*
\def\SEC{References}
\bibliographystyle{unsrt}

\begin{thebibliography}{1}

\bibitem{SW000}
M.~Shub and A.~Wilkinson.
\newblock Pathological foliations and removable zero exponents.
\newblock {\em Inventiones Mathematicae}, 139:495--508, 2000.

\bibitem{RW001}
D.~Ruelle and A.~Wilkinson.
\newblock {Absolutely singular dynamical foliations}.
\newblock {\em Communications in Mathematical Physics}, 219:481--–487, 2001.

\bibitem{Mi997}
J.~Milnor.
\newblock {Fubini foiled: Katok's paradoxical example in measure theory}.
\newblock {\em Mathematical Intellegencer}, 10:31--32, 1997.

\bibitem{BKOVZ002}
S.~Boccaletti, J.~Kurths, G.~Osipov, D.L. Valladares, and C.S. Zhou.
\newblock The synchronization of chaotic systems.
\newblock {\em Physics Reports}, 366:1--101, 2002.

\bibitem{Ru986}
D.~Ruelle.
\newblock Resonances of chaotic dynamical systems.
\newblock {\em Physical Review Letters}, 56:405--407, 1986.

\bibitem{GBG004}
G.~Gallavotti, F.~Bonetto, and G.~Gentile.
\newblock {\em Aspects of the ergodic, qualitative and statistical theory of
  motion}.
\newblock Springer Verlag, Berlin, 2004.

\end{thebibliography}

\end{document}